\documentclass[twocolumn]{aastex701}

\usepackage{longtable}
\usepackage{comment}
\usepackage{tabu}
\usepackage{amsmath,amssymb,graphicx,hyperref}

\newcommand{\nbs}{\nobreak\hspace{0.15em}}

\newcommand{\teq}{\ensuremath{T_{\rm eq}}}

\newcommand{\vsini}{\ensuremath{v \sin{i}}}
\newcommand{\kms}{\ensuremath{{\rm km\,s^{-1}}}}
\newcommand{\ms}{\ensuremath{{\rm m\,s^{-1}}}}
\newcommand{\mjup}{\ensuremath{{M_{\rm Jup}}}}
\newcommand{\mnep}{\ensuremath{{M_{\rm Nep}}}}
\newcommand{\msat}{\ensuremath{{M_{\rm Sat}}}}
\newcommand{\mearth}{\ensuremath{M_{\rm \oplus}}}
\newcommand{\mpl}{\ensuremath{{M_{\rm pl}}}}
\newcommand{\rjup}{\ensuremath{R_{\rm Jup}}}
\newcommand{\rnep}{\ensuremath{R_{\rm Nep}}}
\newcommand{\rsat}{\ensuremath{R_{\rm Sat}}}
\newcommand{\rpl}{\ensuremath{R_{\rm pl}}}
\newcommand{\rstar}{\ensuremath{R_{\rm \star}}}
\newcommand{\mstar}{\ensuremath{M_{\rm \star}}}

\newcommand{\rsun}{\ensuremath{R_{\rm \odot}}}
\newcommand{\msun}{\ensuremath{M_{\rm \odot}}}

\newcommand{\rhopl}{\ensuremath{{\rho_{\rm pl}}}}

\newcommand{\stnameA}{NGTS-11}
\newcommand{\plnameA}{NGTS-11\nbs c}
\newcommand{\plnameB}{NGTS-11\nbs b}

\newcommand{\altplnameB}{TOI-1847\nbs b}

\newcommand{\PA}{\ensuremath{12.77264 \pm 0.000017}} 
\newcommand{\tcA}{\ensuremath{2459117.83838 \pm 0.00058}} 
\newcommand{\pA}{\ensuremath{0.05297 \pm 0.00065}} 
\newcommand{\bA}{\ensuremath{0.808 \pm 0.015}} 
\newcommand{\KA}{\ensuremath{6.4^{+1.6}_{-1.5}}} 

\newcommand{\PB}{\ensuremath{35.455882 \pm 0.000039}} 
\newcommand{\tcB}{\ensuremath{2459135.28040 \pm 0.00055}} 
\newcommand{\pB}{\ensuremath{0.1005 \pm 0.0011}} 
\newcommand{\bB}{\ensuremath{0.846 \pm 0.014}} 
\newcommand{\KB}{\ensuremath{12.9^{+1.7}_{-1.8}}} 

\newcommand{\rhostA}{\ensuremath{1920 \pm 170}} 

\newcommand{\qaTESSTRAPPIST}{\ensuremath{0.11 \pm 0.09}} 
\newcommand{\qbTESSTRAPPIST}{\ensuremath{0.41 \pm 0.31}} 
\newcommand{\qaNGTS}{\ensuremath{0.26 \pm 0.13}} 
\newcommand{\qbNGTS}{\ensuremath{0.18 \pm 0.15}} 
\newcommand{\qaMEarthULMTPEST}{\ensuremath{0.28 \pm 0.17}} 
\newcommand{\qaLCOKeplerCamSAAO}{\ensuremath{0.25 \pm 0.11}} 

\newcommand{\mfluxTESSA}{\ensuremath{-0.00002	\pm	0.00005}} 
\newcommand{\mfluxTESSB}{\ensuremath{-0.00003	\pm	0.00005}} 
\newcommand{\mfluxNGTS}{\ensuremath{0.00002	\pm	0.00003}} 
\newcommand{\mfluxMEarth}{\ensuremath{-0.00002	\pm	0.00008}} 
\newcommand{\mfluxLCO}{\ensuremath{-0.00006	\pm	0.00006}} 
\newcommand{\mfluxKeplerCam}{\ensuremath{0.0000	\pm	0.0002}} 
\newcommand{\mfluxULMT}{\ensuremath{-0.0001	\pm	0.0003}} 
\newcommand{\mfluxTRAPPIST}{\ensuremath{-0.0001	\pm	0.0002}} 
\newcommand{\mfluxPEST}{\ensuremath{-0.0001	\pm	0.0004}} 
\newcommand{\mfluxSPECULOOS}{\ensuremath{0.0003	\pm	0.0002}} 

\newcommand{\sigmaTESSA}{\ensuremath{{0	^{+	10	}_{-	0	}}}} 
\newcommand{\sigmaTESSB}{\ensuremath{{2	^{+	150	}_{-	2	}}}} 
\newcommand{\sigmaNGTS}{\ensuremath{773 \pm 36}} 
\newcommand{\sigmaMEarth}{\ensuremath{{972 \pm 86}}} 
\newcommand{\sigmaLCO}{\ensuremath{832 \pm 54	}} 
\newcommand{\sigmaKeplerCam}{\ensuremath{2660 \pm 120 }} 
\newcommand{\sigmaULMT}{\ensuremath{{0	^{+	39	}_{-	0	}}}} 
\newcommand{\sigmaTRAPPIST}{\ensuremath{{2070 \pm 170 }}} 
\newcommand{\sigmaPEST}{\ensuremath{0	^{+	45	}_{-	0	}}} 
\newcommand{\sigmaSPECULOOS}{\ensuremath{0	^{+	45	}_{-	0	}}} 
\newcommand{\sigmaSAAO}{\ensuremath{1	^{+	51	}_{-	0	}}} 

\newcommand{\muFEROSA}{\ensuremath{21\,384.0 \pm 8.2}} 
\newcommand{\sigmawFEROSA}{\ensuremath{24.4 \pm 5.6}} 
\newcommand{\muHARPSA}{\ensuremath{21\,400.6 \pm 2.0}} 
\newcommand{\dmuA}{\ensuremath{0.072 \pm 0.012}} 
\newcommand{\ddmuA}{\ensuremath{0.000170 \pm 0.000019}} 

\newcommand{\mpA}{\ensuremath{0.064^{+0.019}_{-0.013} }} 
\newcommand{\mpAnep}{\ensuremath{1.2^{+0.3}_{-0.2}}} 
\newcommand{\mpAearth}{\ensuremath{20^{+6}_{-4}}} 
\newcommand{\rpA}{\ensuremath{0.429 \pm 0.009}} 
\newcommand{\rpAnep}{\ensuremath{1.24 \pm 0.03}} 
\newcommand{\rhoA}{\ensuremath{1040 \pm 260}} 
\newcommand{\arstarA}{\ensuremath{25.5 \pm 0.7}} 
\newcommand{\smaA}{\ensuremath{0.099 \pm 0.003}} 
\newcommand{\smaAsimp}{\ensuremath{0.099}} 
\newcommand{\incA}{\ensuremath{88.21 \pm 0.09}} 
\newcommand{\tqA}{\ensuremath{594 \pm 13}} 
\newcommand{\eccA}{\ensuremath{0.09}^{+0.09}_{-0.06}}

\newcommand{\mpB}{\ensuremath{0.189 \pm 0.026}} 
\newcommand{\mpBsat}{\ensuremath{0.63 \pm 0.09}} 
\newcommand{\mpBsatOld}{\ensuremath{1.2 \pm 0.3}} 
\newcommand{\mpBearth}{\ensuremath{60 \pm 8}} 
\newcommand{\rpB}{\ensuremath{0.81 \pm 0.02}} 
\newcommand{\rpBsat}{\ensuremath{0.97 \pm 0.02}} 
\newcommand{\rhoB}{\ensuremath{423 \pm 65}} 
\newcommand{\arstarB}{\ensuremath{50 \pm 2}} 
\newcommand{\smaB}{\ensuremath{0.195 \pm 0.006}} 
\newcommand{\smaBsimp}{\ensuremath{0.195}} 
\newcommand{\incB}{\ensuremath{89.05 \pm 0.04}} 
\newcommand{\tqB}{\ensuremath{423 \pm 9}} 
\newcommand{\eccB}{\ensuremath{0.08^{+0.06}_{-0.05}}}

\submitjournal{ApJ}

\shorttitle{\plnameA: a transiting warm Neptune interior to a warm Saturn}
\shortauthors{D. R. Anderson et al.}

\begin{document}

\title{\plnameA: a transiting Neptune-mass planet interior to the warm Saturn \plnameB}

\correspondingauthor{David R. Anderson}

\author[0000-0001-7416-7522]{David R. Anderson}
\affiliation{Department of Physics, University of Warwick, Gibbet Hill Road, Coventry CV4 7AL, UK}
\affiliation{Instituto de Astronomía, Universidad Católica del Norte, Angamos 0610, 1270709, Antofagasta, Chile}
\email[show]{david.anderson@ucn.cl}

\author[0000-0002-1896-2377]{José I. Vinés}
\affiliation{Instituto de Astronomía, Universidad Católica del Norte, Angamos 0610, 1270709, Antofagasta, Chile}
\email{jose.vines.l@gmail.com}

\author[0000-0002-2135-9018]{Katharine Hesse}
\affiliation{Department of Physics and Kavli Institute for Astrophysics and Space Research, Massachusetts Institute of Technology, Cambridge, MA 02139, USA}
\email{khesse@mit.edu}

\author[0000-0002-5254-2499]{Louise Dyregaard Nielsen}
\affiliation{European Southern Observatory, Karl-Schwarzschild-Straße 2, 85748 Garching bei M\"unchen, Germany}
\email{louise.nielsen@lmu.de}

\author[0000-0002-9158-7315]{Rafael Brahm}
\affiliation{Facultad de Ingeniería y Ciencias, Universidad Adolfo Ibáñez, Av. Diagonal las Torres 2640, Peñalolén, Santiago, Chile}
\affiliation{Millennium Institute for Astrophysics, Chile}
\affiliation{Data Observatory Foundation, Chile}
\email{rafael.brahm@uai.cl}

\author[0000-0002-7927-9555]{Maximiliano Moyano}
\affiliation{Instituto de Astronomía, Universidad Católica del Norte, Angamos 0610, 1270709, Antofagasta, Chile}
\email{mmoyano@ucn.cl}

\author[0000-0003-1452-2240]{Peter J. Wheatley}
\affiliation{Centre for Exoplanets and Habitability, University of Warwick, Gibbet Hill Road, Coventry CV4 7AL, UK}
\affiliation{Department of Physics, University of Warwick, Gibbet Hill Road, Coventry CV4 7AL, UK}
\email{p.j.wheatley@warwick.ac.uk}

\author[0000-0003-1464-9276]{Khalid Barkaoui}
\affiliation{Astrobiology Research Unit, Université de Liège, 19C Allée du 6 Août, 4000 Liège, Belgium}
\affiliation{Department of Earth, Atmospheric and Planetary Science, Massachusetts Institute of Technology, 77 Massachusetts Avenue,\\ Cambridge, MA 02139, USA}
\affiliation{Instituto de Astrofísica de Canarias (IAC), Calle Vía Láctea s/n, 38200, La Laguna, Tenerife, Spain}
\email{khalid.barkaoui@uliege.be}

\author[0000-0001-6637-5401]{Allyson Bieryla}
\affiliation{Center for Astrophysics \textbar \ Harvard \& Smithsonian, 60 Garden St, Cambridge, MA 02138, USA}
\email{abieryla@cfa.harvard.edu}

\author[0000-0003-0684-7803]{Matthew R. Burleigh}
\affiliation{Centre for Exoplanet Research, School of Physics and Astronomy, University of Leicester, Leicester LE1 7RH, UK}
\email{mrb1@le.ac.uk}

\author[0000-0001-5383-9393]{Ryan Cloutier}
\affiliation{Center for Astrophysics \textbar \ Harvard \& Smithsonian, 60 Garden St, Cambridge, MA 02138, USA}
\affiliation{Department of Physics \& Astronomy, McMaster University, 1280 Main St W, Hamilton, ON, L8S 4L8, Canada}
\email{ryan.cloutier@mcmaster.ca}

\author[0000-0001-6588-9574]{Karen A. Collins}
\affiliation{Center for Astrophysics \textbar \ Harvard \& Smithsonian, 60 Garden St, Cambridge, MA 02138, USA}
\email{karen.collins@cfa.harvard.edu}

\author[0000-0002-5674-2404]{Phil Evans}
\affiliation{El Sauce Observatory, Coquimbo Province, Chile}
\email{phil@astrofizz.com}

\author[0000-0002-2532-2853]{Steve B. Howell}
\affiliation{NASA Ames Research Center, Moffett Field, CA 94035}
\email{steve.b.howell@nasa.gov}

\author[0000-0003-0497-2651]{John Kielkopf}
\affiliation{Department of Physics and Astronomy, University of Louisville, Louisville, KY 40292, USA}
\email{kielkopf@louisville.edu}

\author[0000-0003-0828-6368]{Pablo Lewin}
\affiliation{The Maury Lewin Astronomical Observatory, Glendora, CA 91741, USA}
\email{pablotwa@earthlink.net}

\author[0000-0001-8227-1020]{Richard P. Schwarz}
\affiliation{Center for Astrophysics \textbar \ Harvard \& Smithsonian, 60 Garden St, Cambridge, MA 02138, USA}
\email{rpschwarz@comcast.net}

\author[0000-0002-1836-3120]{Avi Shporer}
\affiliation{Department of Physics and Kavli Institute for Astrophysics and Space Research, Massachusetts Institute of Technology, Cambridge, MA 02139, USA}
\email{shporer@mit.edu}

\author[0000-0001-5603-6895]{Thiam-Guan Tan}
\affiliation{Perth Exoplanet Survey Telescope, Perth, Western Australia, Australia}
\email{tgtan@bigpond.net.au}

\author[0009-0008-2214-5039]{Mathilde Timmermans}
\affiliation{Astrobiology Research Unit, Université de Liège, 19C Allée du 6 Août, 4000 Liège, Belgium}
\email{mathilde.timmermans@uliege.be}

\author[0000-0002-5510-8751]{Amaury H. M. J. Triaud}
\affiliation{School of Physics \& Astronomy, University of Birmingham, Edgbaston, Birmingham B15 2TT, United Kingdom}
\email{a.triaud@bham.ac.uk}

\author[0000-0002-0619-7639]{Carl Ziegler}
\affiliation{Department of Physics, Engineering and Astronomy, Stephen F. Austin State University, 1936 North St, Nacogdoches, TX 75962, USA}
\email{Carl.Ziegler@sfasu.edu}

\author[0009-0004-7473-4573]{Ioannis Apergis}
\affiliation{Centre for Exoplanets and Habitability, University of Warwick, Gibbet Hill Road, Coventry CV4 7AL, UK}
\affiliation{Department of Physics, University of Warwick, Gibbet Hill Road, Coventry CV4 7AL, UK}
\email{ioannis.apergis@warwick.ac.uk}

\author[0000-0002-5080-4117]{David J. Armstrong}
\affiliation{Centre for Exoplanets and Habitability, University of Warwick, Gibbet Hill Road, Coventry CV4 7AL, UK}
\affiliation{Department of Physics, University of Warwick, Gibbet Hill Road, Coventry CV4 7AL, UK}
\email{d.j.armstrong@warwick.ac.uk}

\author[0000-0002-5619-2502]{Douglas R. Alves}
\affiliation{Departamento de Astronomía, Universidad de Chile, Casilla 36-D, Santiago, Chile}
\affiliation{Centro de Astrof\'isica y Tecnolog\'ias Afines (CATA), Casilla 36-D, Santiago, Chile}
\email{dalves@das.uchile.cl}

\author[0000-0001-6023-1335]{Daniel Bayliss}
\affiliation{Centre for Exoplanets and Habitability, University of Warwick, Gibbet Hill Road, Coventry CV4 7AL, UK}
\email{d.bayliss@warwick.ac.uk}

\author[0000-0002-7613-393X]{Francois Bouchy}
\affiliation{Geneva Observatory, University of Geneva, Chemin Pegasi 51, 1290 Versoix, Switzerland}
\email{francois.bouchy@unige.ch}

\author[0000-0003-2478-0120]{Sarah L. Casewell}
\affiliation{Centre for Exoplanet Research, School of Physics and Astronomy, University of Leicester, Leicester LE1 7RH, UK}
\email{slc25@le.ac.uk}

\author[0000-0003-0061-5446]{Alexander Chaushev}
\affiliation{University of California Irvine, Department of Physics and Astronomy, Irvine, California, United States}
\email{a.chaushev@uci.edu}

\author[0009-0000-5659-9006]{Benjamin D. R. Davies}
\affiliation{Centre for Exoplanets and Habitability, University of Warwick, Gibbet Hill Road, Coventry CV4 7AL, UK}
\affiliation{Department of Physics, University of Warwick, Gibbet Hill Road, Coventry CV4 7AL, UK}
\email{ben.d.r.davies@warwick.ac.uk}

\author[0000-0002-6939-9211]{Tansu Daylan}
\affiliation{Department of Physics and McDonnell Center for the Space Sciences, Washington University, St. Louis, MO 63130, USA}
\email{tansu@wustl.edu}

\author[0000-0002-7008-6888]{Elsa Ducrot}
\affiliation{LESIA, Observatoire de Paris, CNRS, Universit\'e Paris Diderot, Universit\'e Pierre et Marie Curie, 5 place Jules Janssen, 92190 Meudon, France}
\email{mailto:elsa.ducrot@obspm.fr}

\author[0000-0003-3986-0297]{Mourad Ghachoui}
\affiliation{Astrobiology Research Unit, Université de Liège, 19C Allée du 6 Août, 4000 Liège, Belgium}
\affiliation{Oukaimeden Observatory, High Energy Physics and Astrophysics Laboratory, Faculty of sciences Semlalia, Cadi Ayyad University, Marrakech, Morocco}
\email{mr.ghachoui@gmail.com}

\author[0000-0002-4259-0155]{Samuel Gill}
\affiliation{Centre for Exoplanets and Habitability, University of Warwick, Gibbet Hill Road, Coventry CV4 7AL, UK}
\affiliation{Department of Physics, University of Warwick, Gibbet Hill Road, Coventry CV4 7AL, UK}
\email{samuel.gill@warwick.ac.uk}

\author[0000-0003-2851-3070]{Edward Gillen}
\affiliation{Astronomy Unit, Queen Mary University of London, Mile End Road, London E1 4NS, UK}
\email{e.gillen@qmul.ac.uk}

\author[0000-0003-1462-7739]{Michaël Gillon}
\affiliation{Astrobiology Research Unit, Université de Liège, 19C Allée du 6 Août, 4000 Liège, Belgium}
\email{michael.gillon@uliege.be}

\author[0000-0002-3164-9086]{Maximilian~N.~G\"unther}
\affiliation{European Space Agency (ESA), European Space Research and Technology Centre (ESTEC), Keplerlaan 1, 2201 AZ Noordwijk, The Netherlands}
\email{maximilian.guenther@esa.int}

\author[0000-0002-1493-300X]{Thomas Henning}
\affiliation{Max-Planck-Institut f\"ur Astronomie, K\"onigstuhl 17, 69117 Heidelberg, Germany}
\email{henning@mpia-hd.mpg.de}

\author[0000-0002-5945-7975]{Melissa Hobson}
\affiliation{Max-Planck-Institut f\"ur Astronomie, K\"onigstuhl 17, 69117 Heidelberg, Germany}
\email{melihobson@gmail.com}

\author[0000-0003-1728-0304]{Keith Horne}
\affiliation{SUPA School of Physics and Astronomy, University of St Andrews, St Andrews, KY16 9SS, Scotland, UK}
\email{kdh1@st-andrews.ac.uk}

\author[0000-0001-8923-488X]{Emmanuël Jehin}
\affiliation{Space Sciences, Technologies and Astrophysics Research (STAR) Institute, Université de Liège, Allée du 6 Août 19C, 4000 Liège, Belgium}
\email{ejehin@uliege.be}

\author[0000-0003-2733-8725]{James S. Jenkins}
\affiliation{Instituto de Estudios Astrof\'isicos, Facultad de Ingenier\'ia y Ciencias, Universidad Diego Portales, Av. Ej\'ercito 441, Santiago, Chile}
\affiliation{Centro de Astrof\'isica y Tecnolog\'ias Afines (CATA), Casilla 36-D, Santiago, Chile}
\email{james.jenkins@mail.udp.cl}

\author[0000-0002-5389-3944]{Andres Jordan}
\affiliation{Facultad de Ingeniería y Ciencias, Universidad Adolfo Ibáñez, Av. Diagonal las Torres 2640, Peñalolén, Santiago, Chile}
\affiliation{Millennium Institute for Astrophysics, Chile}
\email{andres.jordan@uai.cl}

\author[0000-0001-9269-8060]{Michelle Kunimoto}
\affiliation{Department of Physics and Kavli Institute for Astrophysics and Space Research, Massachusetts Institute of Technology, Cambridge, MA 02139, USA}
\email{mkuni@mit.edu}

\author[0000-0003-1712-6112]{Régis Lachaume}
\affiliation{Centro de Astroingenier\'ia, Instituto de Astrof\'isica, Pontificia Universidad Cat\'olica de Chile, casilla 306, Santiago 22, Chile}
\affiliation{Max-Planck-Institut f\"ur Astronomie, K\"onigstuhl 17, 69117 Heidelberg, Germany}
\email{lachaume@astro.puc.cl}

\author[0000-0001-9699-1459]{Monika Lendl}
\affiliation{Geneva Observatory, University of Geneva, Chemin Pegasi 51, 1290 Versoix, Switzerland}
\email{monika.lendl@unige.ch}

\author[0000-0003-1631-4170]{James McCormac}
\affiliation{Centre for Exoplanets and Habitability, University of Warwick, Gibbet Hill Road, Coventry CV4 7AL, UK}
\affiliation{Department of Physics, University of Warwick, Gibbet Hill Road, Coventry CV4 7AL, UK}
\email{j.j.mccormac@warwick.ac.uk}

\author[0000-0001-9087-1245]{Felipe Murgas}
\affiliation{Instituto de Astrofísica de Canarias (IAC), Calle Vía Láctea s/n, 38200, La Laguna, Tenerife, Spain}
\affiliation{Departamento de Astrof\'isica, Universidad de La Laguna (ULL), E-38206 La Laguna, Tenerife, Spain}
\email{fmurgas@iac.es}

\author[0000-0001-8504-5862]{Catriona Murray}
\affiliation{Department of Astrophysical and Planetary Sciences, University of Colorado Boulder, Boulder, CO 80309, USA}
\email{mailto:Catriona.Murray@colorado.edu}

\author[0000-0002-5899-7750]{Ares Osborn}
\affiliation{Department of Physics \& Astronomy, McMaster University, 1280 Main St W, Hamilton, ON, L8S 4L8, Canada}
\affiliation{Centre for Exoplanets and Habitability, University of Warwick, Gibbet Hill Road, Coventry CV4 7AL, UK}
\affiliation{Department of Physics, University of Warwick, Gibbet Hill Road, Coventry CV4 7AL, UK}
\email{e.osborn@warwick.ac.uk}

\author[0000-0003-1572-7707]{Francisco J. Pozuelos}
\affiliation{Instituto de Astrof\'isica de Andaluc\'ia (IAA-CSIC), Glorieta de la Astronom\'ia s/n, 18008 Granada, Spain}
\email{pozuelos@iaa.es}

\author[0000-0002-3012-0316]{Didier Queloz}
\affiliation{Cavendish Laboratory, JJ Thomson Avenue, Cambridge CB3 0HE, UK}
\affiliation{Department of Physics, ETH Zurich, Wolfgang-Pauli-Strasse 2, CH-8093 Zurich, Switzerland}
\email{mailto:dq212@cam.ac.uk}

\author[0000-0001-8018-0264]{Suman Saha}
\affiliation{Instituto de Estudios Astrof\'isicos, Facultad de Ingenier\'ia y Ciencias, Universidad Diego Portales, Av. Ej\'ercito 441, Santiago, Chile}
\affiliation{Centro de Astrof\'isica y Tecnolog\'ias Afines (CATA), Casilla 36-D, Santiago, Chile}
\email{suman.saha@mail.udp.cl}

\author[0000-0002-2214-9258]{Daniel Sebastian}
\affiliation{School of Physics \& Astronomy, University of Birmingham, Edgbaston, Birmingham B15 2TT, United Kingdom}
\email{mailto:d.sebastian.1@bham.ac.uk}

\author[0000-0002-2386-4341]{Alexis M. S. Smith}
\affiliation{Institute of Planetary Research, German Aerospace Center (DLR), Rutherfordstr. 2, 12489 Berlin, Germany}
\email{alexis.smith@dlr.de}

\author[0000-0001-7576-6236]{Stephane Udry}
\affiliation{Geneva Observatory, University of Geneva, Chemin Pegasi 51, 1290 Versoix, Switzerland}
\email{stephane.udry@unige.ch}

\author[0000-0003-2417-7006]{Solène Ulmer-Moll}
\affiliation{Geneva Observatory, University of Geneva, Chemin Pegasi 51, 1290 Versoix, Switzerland}
\affiliation{Physikalisches Institut, University of Bern, Gesellschaftsstrasse 6, 3012, Bern, Switzerland}
\email{solene.ulmer-moll@unige.ch}

\author[0000-0001-7246-5438]{Andrew Vanderburg}
\affiliation{Department of Physics and Kavli Institute for Astrophysics and Space Research, Massachusetts Institute of Technology, Cambridge, MA 02139, USA}
\email{andrewv@mit.edu}

\author[0000-0001-6604-5533]{Richard G. West}
\affiliation{Centre for Exoplanets and Habitability, University of Warwick, Gibbet Hill Road, Coventry CV4 7AL, UK}
\affiliation{Department of Physics, University of Warwick, Gibbet Hill Road, Coventry CV4 7AL, UK}
\email{richard.west@warwick.ac.uk}

\begin{abstract}
We report the discovery of \plnameA, a transiting warm Neptune ($P \sim 12.8$ d; $\mpl = \mpAnep \mnep$; $\rpl = \rpAnep \rnep$), 
orbiting closer to the K-dwarf \stnameA\ than the previously reported transiting warm Saturn \plnameB\ ($P \sim 35.5$\nbs d), and we report evidence of a third, outer companion. 
We first detected transits of \plnameA\ in TESS light curves and confirmed them with follow-up transits from NGTS and many other ground-based facilities. 
Radial-velocity monitoring with the HARPS and FEROS spectrographs revealed the mass of \plnameA\ and provided evidence, via a long-term trend, for a long-period companion ($P > 2300$\nbs d and $\mpl \sin i > 3.6$\nbs \mjup). 
Taking into account the two additional bodies in our expanded data sets, we find that the mass of \plnameB\ (\mpl\ = \mpBsat\ \msat; \rpl\ = \rpBsat\ \rsat) is lower than previously reported (\mpl\ = \mpBsatOld\ \msat). 
Given their near-circular, compact orbits, \plnameA\ and b are unlikely to have arrived in place via high-eccentricity migration. 
Instead, they likely formed in situ, or they formed farther out and then moved inward through disk migration.
A comparison of \stnameA\ with the eight other known systems hosting multiple well-characterized warm giants reveals that it is most similar to Kepler-56.
Finally, we find that the commonly used 10-day boundary between hot and warm Jupiters appears empirically well supported.
\end{abstract}


\section{Introduction} \label{sec:intro}

The origins of hot Jupiters (orbital period $P < 10$\nbs d) and warm Jupiters (10\nbs d $\le P <$ 100\nbs d) remain open questions, posing challenges for models of planetary migration (see \citealt{2018ARA&A..56..175D} for a review). 
High-eccentricity tidal migration may explain many hot Jupiters, but it likely cannot account for warm Jupiters in near-circular orbits, especially those with inner companions (e.g., \citealt{2015ApJ...808...14M}).
Rather, disk migration (e.g., \citealt{1996Natur.380..606L, 2020ApJ...904..134H}) and in-situ formation (e.g., \citealt{2016ApJ...825...98H, 2020MNRAS.491.1369A}) are more likely pathways. 

\citet{2020ApJ...898L..11G} reported the discovery of \plnameB\ ($\equiv$\,\altplnameB), a transiting Saturn in a 35.46-day orbit around a K-dwarf star (\autoref{tab:stellar}).
It is the first exoplanet to be discovered based on the detection of a single transit in a TESS light curve \citep{2015JATIS...1a4003R}. 
Intensive monitoring with NGTS \citep{2018MNRAS.475.4476W} resulted in the observation of a second transit. The orbital period was subsequently derived from a combination of the transit photometry and radial velocities from the HARPS \citep{2002Msngr.110....9P} and FEROS \citep{1998SPIE.3355..844K} spectrographs, with the radial-velocity data providing the planetary mass.

\begin{deluxetable}{lll}
\tablecaption{Stellar properties of \stnameA\label{tab:stellar}}
\tablewidth{0pt}
\tabletypesize{\scriptsize}
\tablehead{
\colhead{Parameter (Unit)} & \colhead{Value} & \colhead{Reference}
}
\startdata
Designation     & NGTS-11       & \citet{2020ApJ...898L..11G}   \\
\dotfill        & TOI-1847      &  \citet{2020ApJ...898L..11G}        \\
\dotfill        & TIC\,54002556  & \citet{2019AJ....158..138S} \\
\dotfill        & {\scriptsize 2453680078509741056} & \citet{2023AA...674A...1G} \\
R.A. \dotfill (J2000)     & 01:34:05.14   & \citet{2006AJ....131.1163S} \\
Decl. \dotfill (J2000)    & $-$14:25:09.0 & \citet{2006AJ....131.1163S} \\
$\mu_{\rm R.A.}$ \dotfill (mas/yr)  & 11.11 $\pm$ 0.02 & \citet{2023AA...674A...1G} \\
$\mu_{\rm Decl.}$ \dotfill (mas/yr) & 13.88 $\pm$ 0.01 & \citet{2023AA...674A...1G}\\
$\pi$ \dotfill (mas)  & 5.294 $\pm$ 0.018 & \citet{2023AA...674A...1G} \\
\hline
$T$ \dotfill (mag)    & 11.624 $\pm$ 0.006  & \citet{2019AJ....158..138S} \\
$B$ \dotfill (mag)    & 13.31 $\pm$ 0.04  & \citet{2019AJ....158..138S} \\
$V$ \dotfill (mag)    & 12.46 $\pm$ 0.08  & \citet{2019AJ....158..138S} \\
$G$ \dotfill (mag)    & 12.177 $\pm$ 0.003  & \citet{2023AA...674A...1G} \\
$G_{\rm BP}$ \dotfill (mag)    & 12.646 $\pm$ 0.003  & \citet{2023AA...674A...1G} \\
$G_{\rm RP}$ \dotfill (mag)    & 11.559 $\pm$ 0.004  & \citet{2023AA...674A...1G} \\
$J$ \dotfill (mag)    & 10.86 $\pm$ 0.03 & \citet{2006AJ....131.1163S}\\
$H$ \dotfill (mag)    & 10.40 $\pm$ 0.02 & \citet{2006AJ....131.1163S}\\
$K$ \dotfill (mag)    & 10.32 $\pm$ 0.03 & \citet{2006AJ....131.1163S}\\
\hline
     $T_{\rm eff}$ \dotfill (K) & $5050 \pm 80$ &  \citet{2020ApJ...898L..11G} \\
     {[Fe/H]} \dotfill (dex) & $0.22 \pm 0.08$  &  \citet{2020ApJ...898L..11G} \\
     $\log g$ \dotfill (dex) & $4.5 \pm 0.1$ &  \citet{2020ApJ...898L..11G} \\
     $v \sin i$ \dotfill (\kms) & $1.1 \pm 0.8$ & \citet{2020ApJ...898L..11G}\\
     \mstar \dotfill (\msun) & $0.862 \pm 0.028$ & \citet{2020ApJ...898L..11G} \\
     \rstar \dotfill (\rsun) & $0.832 \pm 0.013$ & \citet{2020ApJ...898L..11G} \\
     \hline
\enddata
\end{deluxetable}

A second, shallower transit signal was identified in the same TESS light curve, with a period of 12.77\nbs d. This was confirmed by subsequent transit detections in long-baseline light curves from both TESS (as also noted by \citealt{2022ApJS..259...62I}) and NGTS, and by many ground-based imagers in targeted observations.  
In this paper, we present new transit light curves for both planets and a significantly expanded set of radial velocities, which confirm the planetary origin of the 12.77-day signal, revise the mass estimate of \plnameB, and reveal evidence for an additional, longer-period companion.

\section{Observations} \label{sec:obs}

\subsection{Photometry from imaging} \label{sec:phot}
The first transit of \plnameB, with a depth of $\sim$1~percent, was detected by TESS \citep{2015JATIS...1a4003R} during its Sector 3 campaign (2018 Sep--Oct). 
Based on this detection, we intensively monitored \stnameA\ with NGTS \citep{2018MNRAS.475.4476W}, which detected a second transit on 2019 Oct 24 \citep{2020ApJ...898L..11G}.
Likewise, \stnameA\ was selected for a 20-s cadence light curve in Sector 30 (2020 Sep--Oct; Guest Investigator Program ID: G03092, PI: N. Espinoza), in which a third transit of \plnameB\ was observed, confirming the orbital period of 35.5\nbs d reported by \citet{2020ApJ...898L..11G}.

We identified \plnameA\ in the Sector~30 TESS light curve from two much shallower transits, each $\sim$0.3~percent deep and separated by $\sim$12.8\nbs d.
In hindsight, two complete transits of \plnameA\ were evident in TESS Sector 3, and partial transits are evident in the light curve obtained by NGTS during its monitoring campaign. 
These were initially overlooked owing to the low signal-to-noise of the transits.

Our goal was to confirm the shallow transit signal and refine the transit shape and ephemeris.
On 2020 Nov 28, we observed a transit of \plnameA\ with multiple facilities across the Americas. 
That effort was coordinated through the ground-based time-series photometry subgroup (SG1) of the TESS Follow-up Observing Program (TFOP).

\begin{figure*}
\begin{minipage}{0.5\textwidth}
\includegraphics[width=90mm]{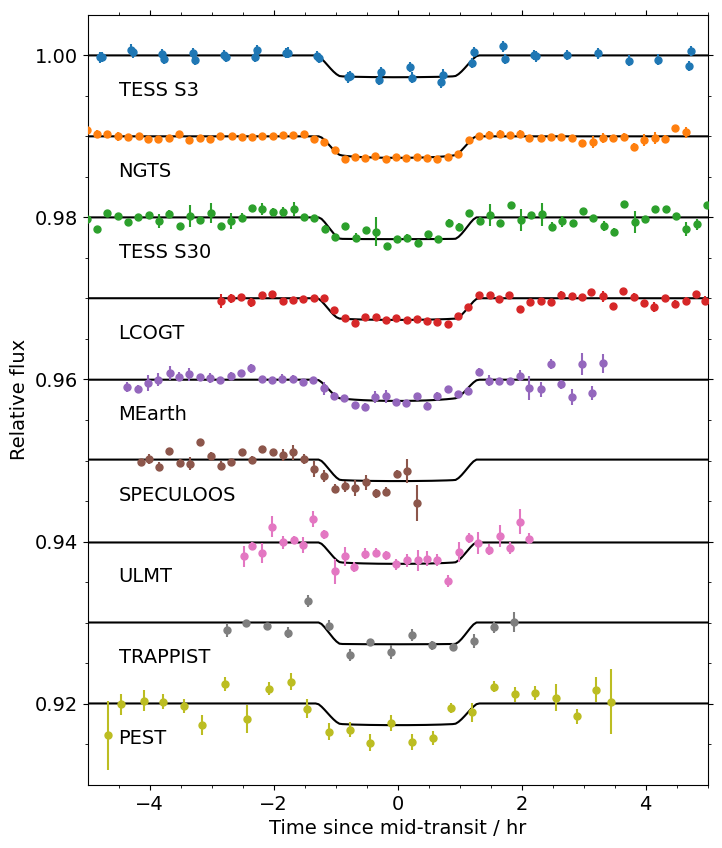}
\end{minipage}
\begin{minipage}{0.5\textwidth}
\includegraphics[width=90mm]{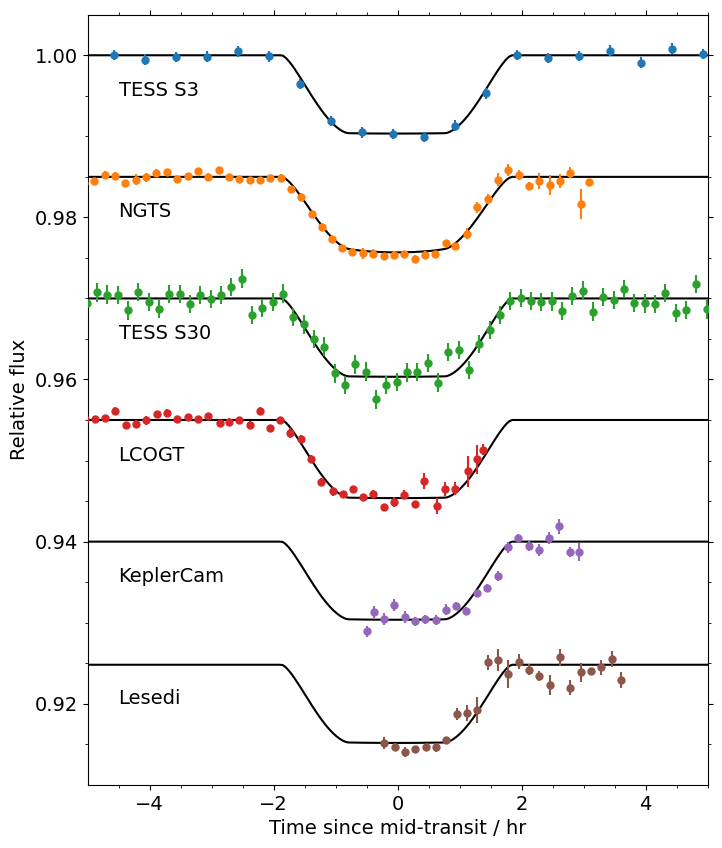}
\end{minipage}
\caption{Transit light curves of \plnameA\ (left panel) and \plnameB\ (right panel), shown offset in relative flux for clarity and plotted to the same scale for comparison. 
The data were phase-binned to 10\nbs min, except for TESS Sector 3 (30\nbs min) and TRAPPIST and PEST (20\nbs min for both). 
The error bars represent formal uncertainties and do not include the photometric jitter terms (\autoref{tab:props}).
The transit model (\autoref{sec:anal}), including limb darkening, is shown for each case.
\label{fig:phot}}
\end{figure*}

Subsequently, we targeted transits of both \plnameA\ and b with both NGTS and SG1 facilities.
A summary of the photometric observations is provided in \autoref{tab:phot}, and the corresponding light curves are shown in \autoref{fig:phot}.

\begin{deluxetable*}{lccccccc}
\tabletypesize{\scriptsize}
\tablecaption{Summary of photometric observations during planetary transits\label{tab:phot}}
\tablehead{
\colhead{Date}	&	\colhead{Facility}	&	\colhead{Reference}	&	\colhead{Apertures}	&	\colhead{Filter}	&	\colhead{$T_{\rm exp}$}	&	\colhead{$T_{\rm c}$ (BJD TDB)}	&	\colhead{$\Delta T_{\rm c}$ (UTC)}\\	
\colhead{\ldots}	&	\colhead{\ldots} & \colhead{\ldots}	&	\colhead{([N $\times$] {\rm cm})}	&	\colhead{\ldots}	&	\colhead{(s)}	&	\colhead{(days)}	&	\colhead{(days)}	
}
\startdata
\noalign{\medskip}			
{\bf \plnameA}															\\
2018-09-28	&	TESS	&	1	&	10.5	&	TESS	&	1800	&	2458389.7961	&	0.0019	\\
2018-10-10	&	TESS	&	1	&	10.5	&	TESS	&	1800	&	2458402.5723	&	0.0035	\\
2019-08-26	&	NGTS	&	2	&	20	&	NGTS	&	10	&	2458721.8854	&	0.0053	\\
2019-10-29	&	NGTS	&	2	&	20	&	NGTS	&	10	&	2458785.7530	&	0.0032	\\
2020-09-25	&	TESS	&	1	&	10.5	&	TESS	&	120	&	2459117.8297	&	0.0035	\\
2020-10-08	&	TESS	&	1	&	10.5	&	TESS	&	120	&	2459130.6076	&	0.0041	\\
2020-11-28	&	ElSauce	&	3	&	36	&	$R_{\rm C}$	&	120	&	2459181.7151	&	0.0223	\\
2020-11-28	&	LCOGT-McDonald	&	4	&	100	&	$i^\prime$	&	40	&	2459181.7016	&	0.0037	\\
2020-11-28	&	MEarth-North	&	5	&	$6 \times 40$	&	RG715	&	28	&	2459181.7025	&	0.0018	\\
2020-11-28	&	MEarth-South	&	5	&	$7 \times 40$	&	RG715	&	49	&	2459181.7013	&	0.0015	\\
2020-11-28	&	MLO-Lewin	&	\ldots	&	36	&	$i^\prime$	&	180	&	2459181.6520	&	0.0274	\\
2020-11-28	&	NGTS	&	2	&	$6 \times 20$	&	NGTS	&	10	&	2459181.7040	&	0.0016	\\
2020-11-28	&	SPECULOOS-Ganymede	&	6	&	100	&	$g^\prime$	&	20	&	2459181.6913	&	0.0041	\\
2020-11-28	&	TRAPPIST	&	7	&	60	&	$I$+$z^\prime$	&	15	&	2459181.7077	&	0.0027	\\
2020-11-28	&	ULMT	&	8	&	60	&	$r^\prime$	&	128	&	2459181.7026	&	0.0028	\\
2021-08-23	&	NGTS	&	2	&	$3 \times 20$	&	NGTS	&	10	&	2459449.9224	&	0.0030	\\
2021-09-05	&	LCOGT-CTIO	&	4	&	100	&	$i^\prime$	&	40	&	2459462.7008	&	0.0012	\\
2021-09-05	&	LCOGT-SAAO	&	4	&	100	&	$i^\prime$	&	40	&	2459475.4759	&	0.0040	\\
2021-09-05	&	NGTS	&	2	&	$3 \times 20$	&	NGTS	&	10	&	2459462.6916	&	0.0039	\\
2021-10-26	&	LCOGT-CTIO	&	4	&	100	&	$i^\prime$	&	40	&	2459513.7929	&	0.0011	\\
2021-10-26	&	LCOGT-McDonald	&	4	&	100	&	$i^\prime$	&	40	&	2459513.7882	&	0.0016	\\
2021-10-26	&	NGTS	&	2	&	$4 \times 20$	&	NGTS	&	10	&	2459513.7904	&	0.0142	\\
2021-12-03	&	PEST	&	9	&	30	&	$R_{\rm C}$	&	120	&	2459552.1153	&	0.0047	\\
2021-12-29	&	NGTS	&	2	&	$4 \times 20$	&	NGTS	&	10	&	2459577.6566	&	0.0091	\\
2022-08-03	&	NGTS	&	2	&	$3 \times 20$	&	NGTS	&	10	&	2459794.7905	&	0.0054	\\
2022-10-06	&	NGTS	&	2	&	$9 \times 20$	&	NGTS	&	10	&	2459858.6514	&	0.0016	\\
2022-10-31	&	PEST	&	9	&	30	&	$r^\prime$	&	120	&	2459884.1837	&	0.0049	\\
\noalign{\medskip}															
{\bf \plnameB}															\\
2018-09-29	&	TESS	&	1	&	10.5	&	TESS	&	1800	&	2458390.7060	&	0.0015	\\
2019-10-24	&	NGTS	&	2	&	20	&	NGTS	&	10	&	2458780.7189	&	0.0015	\\
2020-10-12	&	TESS	&	1	&	10.5	&	TESS	&	120	&	2459135.2801	&	0.0013	\\
2021-01-27	&	LCOGT-CTIO	&	4	&	100	&	$i^\prime$	&	40	&	2459241.6656	&	0.0342	\\
2021-10-02	&	KeplerCam	&	10	&	120	&	$i^\prime$	&	15	&	2459489.8419	&	0.0007	\\
2021-10-02	&	LCOGT-CTIO	&	4	&	100	&	$i^\prime$	&	40	&	2459489.8395	&	0.0016	\\
2021-10-02	&	NGTS	&	2	&	$4 \times 20$	&	NGTS	&	10	&	2459489.8443	&	0.0012	\\
2022-08-17	&	NGTS	&	2	&	$5 \times 20$	&	NGTS	&	10	&	2459808.9482	&	0.0022	\\
2022-10-27	&	NGTS	&	2	&	$4 \times 20$	&	NGTS	&	10	&	2459879.8537	&	0.0017	\\
2022-12-01	&	SAAO/Lesedi	&	\ldots	&	100	&	$i^\prime$	&	10	&	2459915.2973	&	0.0023\\
\noalign{\medskip}
\enddata
\tablerefs{
1. \citet{2015JATIS...1a4003R}.
2. \citet{2018MNRAS.475.4476W}.
3. \citet{2021RMxAC..53...47R}.
4. \citet{2013PASP..125.1031B}.
5. \citet{2008PASP..120..317N}.
6. \citet{2018SPIE10700E..1ID}.
7. \citet{2011Msngr.145....2J}.
8. \url{http://www.astro.louisville.edu/mtlemmon}.
9. \url{http://pestobservatory.com}.
10. \citet{2005AAS...20711010S}.
}
\end{deluxetable*}

\subsection{Radial velocities from spectroscopy} \label{sec:spec}
\begin{figure*}
\centering
\includegraphics[width=180mm]{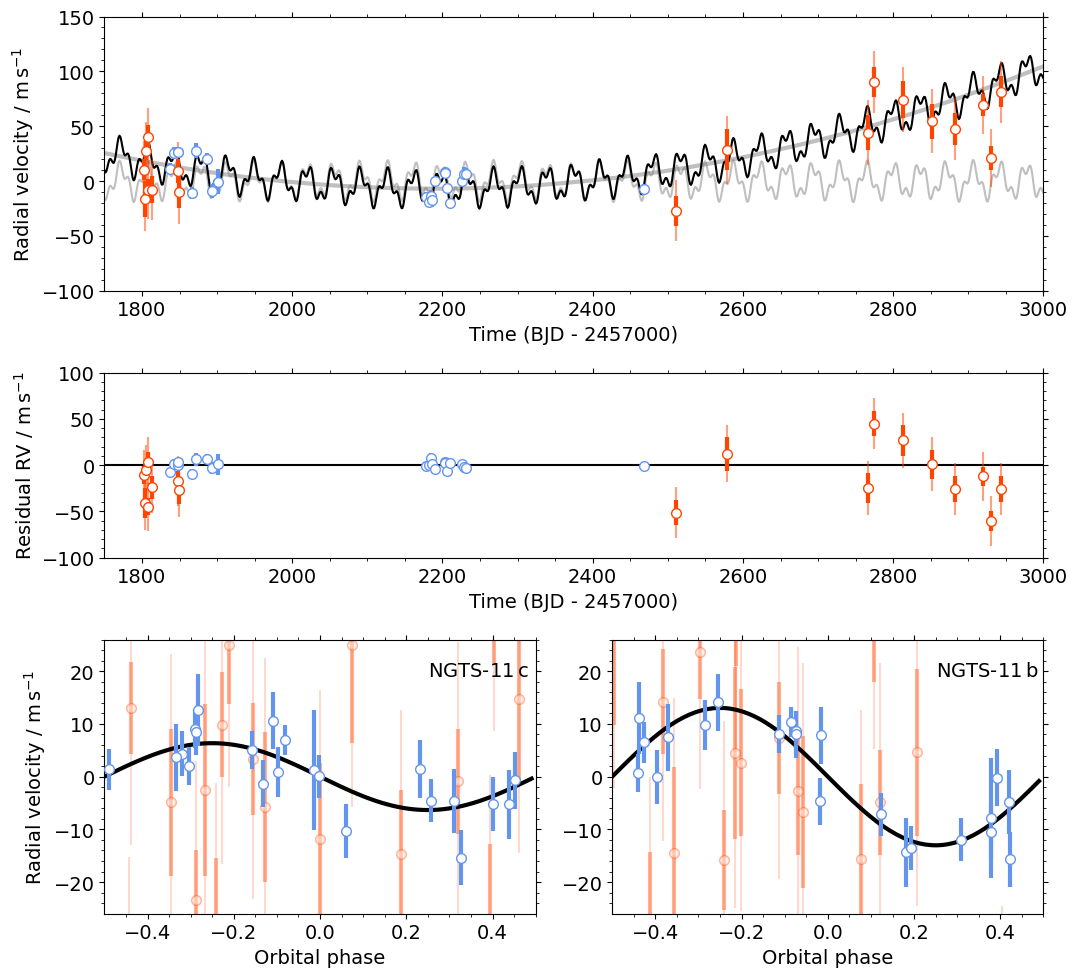}
\caption{{\it Top panel}: RV data from FEROS (red) and HARPS (blue) and the adopted two-planet circular Keplerian model with a quadratic trend (black). The Keplerian signal alone is shown in gray. For the FEROS data, the thicker error bars indicate the formal uncertainties, whereas the thinner error bars show the quadrature sums of the formal uncertainties and the jitter term, $\sigma_{\rm FEROS}$ (\autoref{sec:rv}). 
{\it Middle panel}: Residual RVs after subtracting the best-fitting model. 
{\it Lower-left panel}: Phase-folded RVs for the orbit of \plnameA, after subtraction of the quadratic trend and the Keplerian signal due to \plnameB.
{\it Lower-right panel}: Phase-folded RVs for the orbit of \plnameB, after subtraction of the quadratic trend and the Keplerian signal due to \plnameA.
\label{fig:rvs}}
\end{figure*}

We monitored \stnameA\ with the HARPS \citep{2002Msngr.110....9P}
and FEROS \citep{1999Msngr..95....8K} spectrographs, with typical exposure times of 20\nbs min for FEROS and 30\nbs min for HARPS. 
We acquired 20 FEROS spectra during 2019~Nov--2022~Dec and 21 HARPS spectra during 2019~Dec--2021~Sep.
We extracted radial velocities (RVs) from the HARPS spectra using the techniques described in \citet{1996A&AS..119..373B} and \citet{2002Msngr.110....9P},
and from the FEROS spectra using {\sc ceres} \citep{2017PASP..129c4002B}.
The RVs are listed in \autoref{tab:rvs} and are plotted in \autoref{fig:rvs}. 
RVs from the earliest six FEROS spectra and the earliest nine HARPS spectra had previously been presented in \citet{2020ApJ...898L..11G}.

\begin{deluxetable}{ccccccc}
\tabletypesize{\scriptsize}
\tablecaption{Radial velocities of NGTS-11\label{tab:rvs}}
\tablewidth{0pt}
\tablehead{													
\colhead{Instrument}	&	\colhead{TJD\tablenotemark{a}}	&	\colhead{$\rm RV$\tablenotemark{b}}	&	\colhead{$\Delta \rm RV$}	&	\colhead{FWHM}	&	\colhead{Bis.Sp.}	&	\colhead{SNR}	\\
\ldots	&	(days)	&	(\ms)	&	(\ms)	&	(\kms)	&	(\ms)	&	\ldots	
}										
\startdata
FEROS	&	1802.606589	&	10.5	&	10.0	&	9.254	&	20	&	47	\\
FEROS	&	1803.570007	&	$-$16.3	&	16.2	&	9.511	&	$-$23	&	28	\\
\ldots	&	\ldots	&	\ldots	&	\ldots	&	\ldots	&	\ldots	&	\ldots	\\
HARPS	&	2231.537325	&	6.82	&	4.74	&	7.0291	&	$-$34.2	&	23.6	\\
HARPS	&	2467.827348	&	$-$7.62	&	5.09	&	7.0047	&	$-$29.7	&	22.4	\\
\enddata
\tablenotetext{a}{TESS Julian Date = BJD (TDB) $-$ 2\,457\,000.}
\tablenotetext{b}{A value of 21\,400\,\ms\ was subtracted.}
\end{deluxetable}

\subsection{High-resolution imaging}
\begin{figure*}
\begin{minipage}{0.5\textwidth}
  \includegraphics[width=90mm]{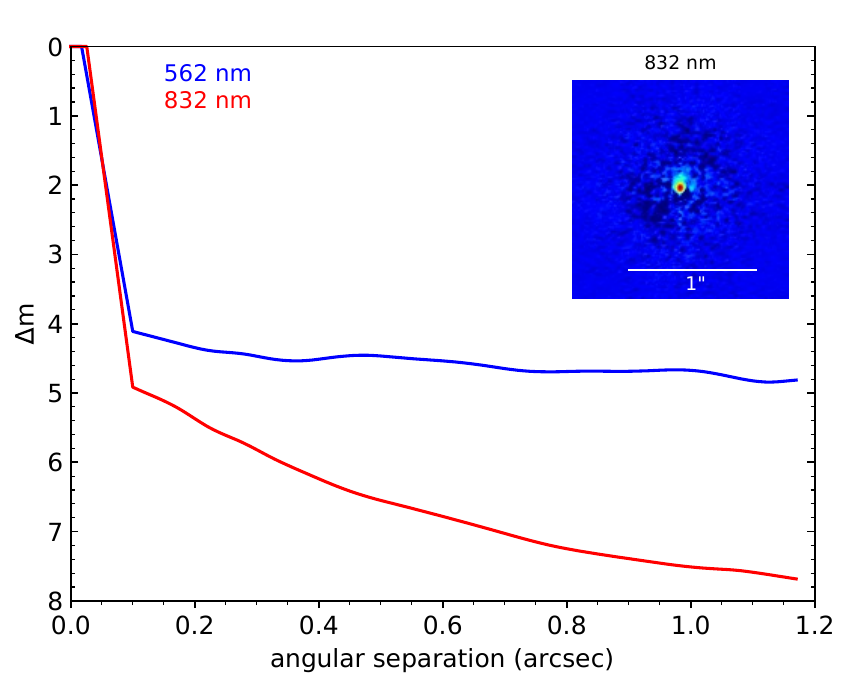}
\end{minipage}
\begin{minipage}{0.5\textwidth}
  \includegraphics[width=90mm]{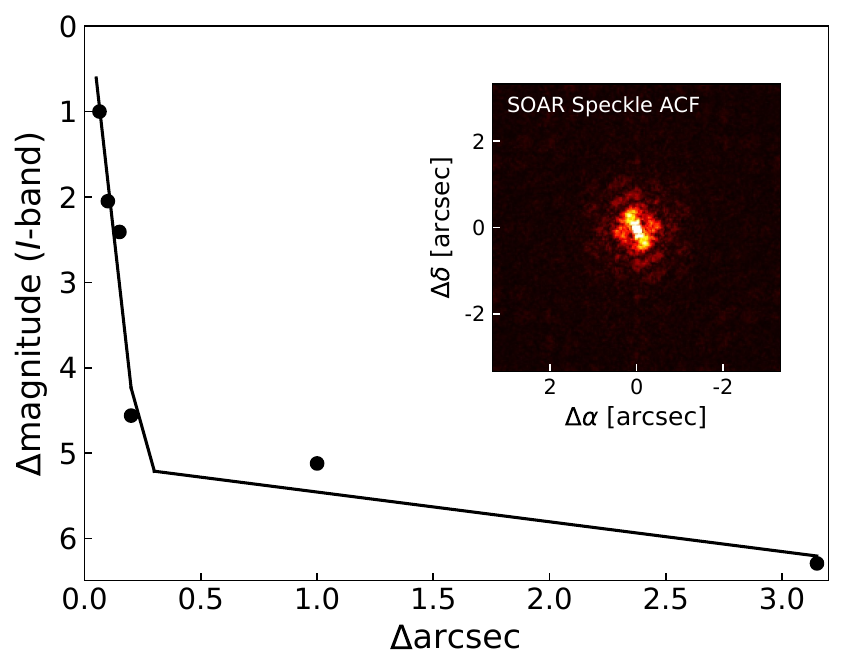}
\end{minipage}
\caption{Speckle imaging of \stnameA\ by 'Alopeke at the Gemini North telescope (left) and HRCam at the Southern Astrophysical Research telescope (right). 
The contrast curve plots show the linear fits to the 5-$\sigma$ contrast curves on either side of 0\farcs1 ('Alopeke) and 0\farcs2 (HRCam). The autocorrelation functions are shown in the insets.
\label{fig:speckle}}
\end{figure*}

We obtained speckle images of \stnameA\ with `Alopeke on Gemini North \citep{2021FrASS...8..138S} on 2020 Aug 10 and with HRCam on SOAR \citep{2018PASP..130c5002T} on 2020 Oct 31. 
The achieved contrasts were $\Delta$6.6\nbs mag at 0.5$^{\prime\prime}$ (`Alopeke) and $\Delta$6.3\nbs mag at 1$^{\prime\prime}$ (HRCam), with no evidence of a stellar companion (\autoref{fig:speckle}).

\section{Analysis} \label{sec:anal}

We modeled the \stnameA\ system with the {\sc juliet} package \citep{2019MNRAS.490.2262E}. The transit light curves were modeled by {\sc batman} \citep{2015PASP..127.1161K} with linear and quadratic limb-darkening parameterization, the radial-velocity variations were modeled with {\sc radvel} \citep{2018PASP..130d4504F}, and the posterior sampling was carried out using the {\sc dynesty} nested sampling algorithm \citep{2020MNRAS.493.3132S}. 
The priors are given in \autoref{tab:props}, along with the median values and the 1-$\sigma$ uncertainties of the posterior distributions. The derived parameters were calculated using the stellar parameters in \autoref{tab:stellar}.

\subsection{Transit light curve modeling}

We obtained a preliminary transit model from an initial {\sc juliet} fit of the TESS-SPOC light curve (Sector 3; 30-min cadence; \citealt{2016SPIE.9913E..3EJ}) and the SPOC light curve (Sector 30; 2-min cadence; \citealt{2020RNAAS...4..201C}). Both were flattened with {\sc Lightkurve} after masking the transits \citep{2018ascl.soft12013L}. The NGTS light curves from 2019, produced by the regular NGTS pipeline, were also included.
The other light curves from ground-based facilities were then fit on a per-night, per-facility basis, with that preliminary transit model fixed (though with transit epoch, $T_{\rm c}$, a free parameter), using the MCMC algorithm of \citet{2007MNRAS.380.1230C}. That included the NGTS light curves from 2020--2022, which had been processed by the quick-look NGTS pipeline. A quadratic temporal trend, as may result from secondary extinction, was fit and removed from the light curve of each telescope, thereby ensuring that each light curve was correctly normalized. 

For computational efficiency, we performed the following steps: 
(i) we cut sections from the TESS light curves with durations of 20\nbs hr ($\sim$8 and $\sim$5 transit durations for \plnameA~and~b, respectively) and centered on the predicted transit times;
(ii) we used the 2-min cadence SPOC light curve from TESS Sector 30, though a 20-s cadence light curve was available;
(iii) we used only those NGTS data taken on transit nights, combined light curves from multiple (1--9) telescopes, and binned in time the combined light curve using a bin width of 2\nbs min;
(iv) we binned the combined MEarth light curve using a 2~min bin width
(v) we binned the TRAPPIST light curve using a bin width of 2\nbs min.

The transit light curves, together with the best-fitting model from the global analysis, are shown in \autoref{fig:phot}.
The \plnameA\ transit light curves taken on 2020 Oct 28 by the Maury Lewin and El~Sauce telescopes (not plotted; \citealt{2021RMxAC..53...47R}) were excluded from the analysis because, with a depth of just 2.5\nbs mmag, the transit proved too shallow to be robustly detected by those 0.36-m telescopes on that particular night.

We calculated a periodogram of the NGTS light curve spanning 2019 Aug 10--Oct 28.
The periodogram showed a broad peak centered on 33\nbs d with a signal amplitude of $\sim$1\nbs mmag, which may correspond to modulation of the light curve caused by stellar rotation and an non-axisymmetric distribution of starspots. 
This is consistent with the rotation period of $\sim$38\nbs d calculated using the \vsini\ and \rstar\ values from \citet{2020ApJ...898L..11G}, though the uncertainties are large. 
In the global analysis, rotational modulation was mitigated through normalization of the individual transit light curves.

\subsection{Radial velocity modeling} \label{sec:rv}

We performed several {\sc juliet} fits to the RV data to determine the appropriate model, with variables of orbital eccentricity and temporal trend, and using the fit to all light curves to place normal priors on the period and the transit epoch. 
Quadratic temporal-trend models are clearly favored over linear or constant models, both when fitting the HARPS data alone and when combining the HARPS and FEROS data.
This suggests the presence of an additional body in the system, on an orbit exterior to that of \plnameB.

Because the evidence is similar for both circular and eccentric orbital models, we adopted the simpler circular orbital model. 
When fitting for eccentricity, the median and 1-$\sigma$ limits are $e = \eccA$ for \plnameA\ and $e = \eccB$ for \plnameB. 

Because we observed scatter in the FEROS RVs in excess of their formal error bars, we added a jitter term, $\sigma_{\rm FEROS}$, in quadrature with the formal errors to achieve a reduced spectroscopic $\chi^2$ of $\sim1$. That proved unnecessary for the HARPS data. 
We excluded two low-S/N FEROS measurements (19 and 23) taken on the nights of 2022 Jun 28 and 2022 Sep 11.

The RVs, together with the best-fitting model from the global analysis, are shown in \autoref{fig:rvs}.

\subsection{A check for transit-timing variations} \label{sec:ttv}
\begin{figure}
\centering
\includegraphics[width=90mm]{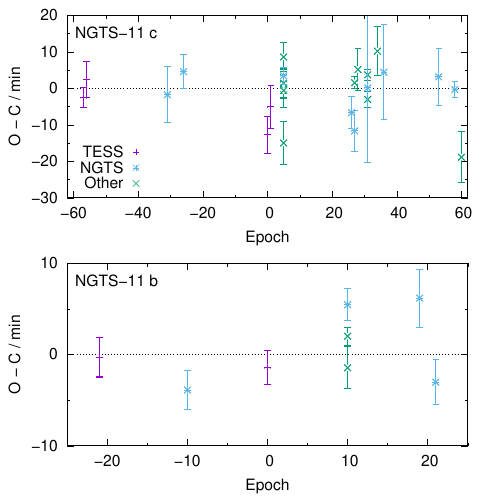}
\caption{
Times of mid-transit (\autoref{tab:phot}) relative to the constant orbital ephemeris of  \autoref{tab:props}, the $T_{\rm c}$ values of which define the zeroth epochs shown here. A few mid-transit times with large uncertainties fall outside of the plotted area.
\label{fig:ttv}}
\end{figure}

We derived individual transit times by fitting each transit light curve on a per-facility, per-epoch basis (\autoref{tab:phot}), fixing the transit shape to that of the global model (\autoref{sec:anal}). 
The results place an upper limit of a few minutes on possible transit-timing variations of the two planets over the observational baseline (\autoref{fig:ttv}). 
The absence of any significant variations in the transit times is unsurprising given that the planets are not near a mean-motion resonance (with a period ratio of $\sim$2.78), appear to have small eccentricities (\autoref{sec:rv}), and have relatively low masses (\autoref{tab:props}).

\section{Discussion} \label{sec:disc}
\begin{figure}
\centering
\includegraphics[width=84mm]{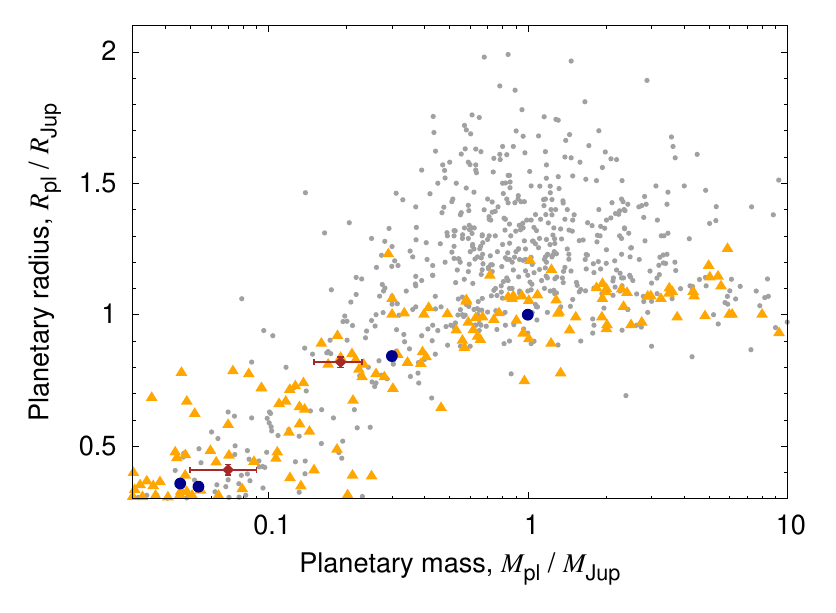}
\caption{
Mass--radius distribution of known exoplanets (gray dots have $P<10$\nbs d and orange triangles have 10\nbs $\le P \le 200$\nbs d), solar system giants (blue circles), and \plnameA\ and~b (brown diamonds, with error bars).
At \mpl\ $\sim 0.2$\nbs \mjup, \plnameB\ approximately marks the mass boundary above which hot Jupiters are often inflated, whereas warm Jupiters generally are not.
Data were retrieved on 2025 Oct 1 from the composite data table of the NASA Exoplanet Archive. 
Planets with masses inferred from a mass--radius relationship were excluded, as were those with imprecise parameters ($\Delta$\mpl/\mpl\ $>$ 0.5 or $\Delta$\rpl/\rpl\ $>$ 0.2). 
\label{fig:m-r}}
\end{figure}

The \stnameA\ system hosts two transiting warm giant planets (\autoref{fig:m-r}) and a third, as-yet uncharacterized body in an outer orbit.
\plnameA\ is a Neptune-mass planet in a 12.8-day orbit ($\mpl =  \mpAnep \mnep \equiv \mpAearth \mearth$; $\rpl = \rpAnep \rnep \equiv \rpA \rjup$; $\teq \sim 594 {\rm K}$ for $A = 0.5$), and thus lies in the Neptunian Savanna \citep{2023A&A...669A..63B,2024A&A...689A.250C}. 
\plnameB\ is a sub-Saturn-mass planet in a 35.5-day orbit ($\mpl\ = \mpBsat\ \msat\ \equiv \mpBearth\ \mearth$; $\rpl\ = \rpBsat\ \rsat\ \equiv \rpB\ \rjup$). 
Based on fewer RVs and with knowledge of only \plnameB, \citet{2020ApJ...898L..11G} derived a higher mass (\mpl\ = \mpBsatOld\ \msat).
More RV data will constrain the mass and period of the outer body, and more photometry will determine whether it transits and, if so, will constrain its size. 
Additional data will also probe for other bodies in the system. 

Considering the orbital separations of \plnameA\ and~b ($a \sim \smaAsimp$\nbs au and $a \sim \smaBsimp$\nbs au, respectively) and their apparently low eccentricities, they seem unlikely to have reached their current locations via a high-eccentricity migration mechanism (e.g., \citealt{2015ApJ...808...14M}).
Instead, the system is likely a product of disk migration (e.g., \citealt{1996Natur.380..606L, 2020ApJ...904..134H}) or in-situ formation (e.g., \citealt{2016ApJ...825...98H, 2020MNRAS.491.1369A}).

\begin{figure}
\centering
\includegraphics[width=84mm]{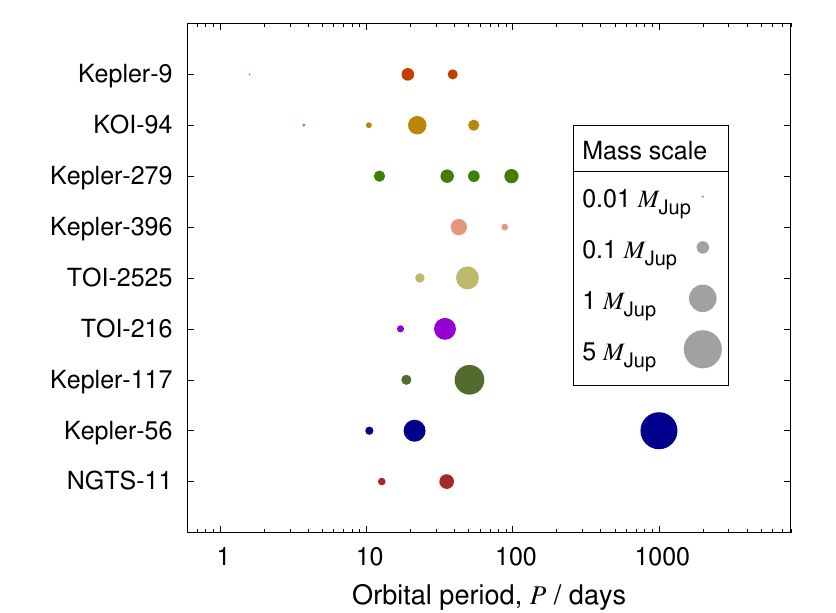}
\caption{
Planetary systems with two or more transiting warm giants satisfying the following constraints: $10 \le P < 100$\nbs d, $0.05 < \mpl < 2 \mjup$, and $0.3 < \rpl < 2 \rjup$. 
The size of each symbol is proportional to the logarithm of the planetary mass. 
Because the mass of Kepler-279\nbs b has not yet been significantly determined, for display purposes, we assumed its mass from a linear interpolation of the masses and radii of Kepler-279\nbs c and~d.  
Data retrieved on 2025 Sep 28 from the composite data table of the NASA Exoplanet Archive. 
\label{fig:wg-pairs}}
\end{figure}

A comparison of \stnameA\ with the eight other known systems hosting multiple well-characterized warm giants reveals that in about half of these systems (5/9), the inner giant is the less massive planet (\autoref{fig:wg-pairs}). 
\stnameA\ joins 
TOI-2525 \citep{2023AJ....165..179T}, 
TOI-216 \citep{2019MNRAS.486.4980K, 2021AJ....161..161D},
Kepler-117 \citep{2014ApJ...784...45R, 2015A&A...573A.124B}, 
and Kepler-56 \citep{2013MNRAS.428.1077S, 2016AJ....152..165O}
in this grouping. 
Conversely, the inner warm giant is more massive in four other of these systems:
Kepler-396 \citep{2014ApJS..210...25X},
Kepler-279 \citep{2014ApJS..210...25X, 2014ApJ...784...45R, 2025AJ....169...90O}, 
KOI-94 \citep{2013ApJ...768...14W}, and 
Kepler-9 \citep{2010Sci...330...51H, 2011ApJ...727...24T, 2019MNRAS.484.3233B}.

Among these systems, \stnameA\ is most similar to Kepler-56. 
Kepler-56 hosts two transiting planets in coplanar orbits that are significantly misaligned with the spin of their host star ($\sim45\deg$; \citealt{2013Sci...342..331H}), as well as an outer super-Jupiter ($\sim6$\nbs \mjup\ in a $\sim$1000-d orbit; \citealt{2016AJ....152..165O}). 
\citet{2017MNRAS.464.1709G} proposed that such misalignments of inner planets can be gently induced by scattering events among unstable outer planets. Therefore, measuring the spin--orbit alignment of \plnameA\ and~b will provide an important test of this formation pathway. 
We predict the amplitude of the Rossiter--McLaughlin effect to be 1\nbs \ms\ for \plnameA\ and 4\nbs \ms\ for \plnameB, signals readily detectable with ESO's VLT/ESPRESSO ($\Delta$RV $\approx$ 2\nbs \ms; \citealt{2021A&A...645A..96P}).

In \autoref{fig:m-r}, we plot \plnameA\ and~b in a mass--radius diagram, together with all the other known well-characterized planets. We use different symbols and colors to distinguish planets in short and longer orbits ($P < 10$\nbs d and 10\nbs  $\le P < $ 100\nbs d, respectively). 

Planets more massive than \plnameB\ (i.e., gas giants) and in sub–10-day orbits (i.e., hot Jupiters) are often inflated, in contrast to gas giants in longer orbits (i.e., warm Jupiters), which are rarely inflated. 
This shows that, even without considering host-star properties, 10\nbs d is a reasonable choice for the boundary between hot Jupiters and warm Jupiters.  

\begin{deluxetable*}{llccc}[!htbp]
\centerwidetable
\tabletypesize{\scriptsize}
\tablecaption{
Prior and posterior parameters of the global analysis of the \stnameA\ system. For the priors: $N(\mu,\sigma)$ represents a normal distribution with mean $\mu$ and standard deviation $\sigma$; $U(a,b)$ represents a uniform distribution between $a$ and $b$; $LU(a,b)$ represents a log-uniform prior defined between $a$ and $b$; and $F(a)$ represents a value fixed at $a$.
The reference epoch for the quadratic trend in the RVs was BJD = 2\,459\,400\,d.
\label{tab:props}
}
\tablehead{
  Fitted parameter & Symbol \dotfill (Unit) & \colhead{Prior} & \plnameA & \plnameB
  }
\startdata
\hline
\hline
\noalign{\smallskip}
Orbital period & $P$ \dotfill (days)   & $N(12.772,0.1)$   & \PA       & \dotfill    \\
\dotfill & \dotfill          & $N(35.456,0.1)$   & \dotfill  & \PB \\
Epoch of mid-transit & $T_{\rm c}$ \dotfill (BJD)&  $N(2459117.839,0.1)$&    \tcA    & \dotfill      \\
\dotfill &  \dotfill          &  $N(2459135.278,0.1)$& \dotfill   & \tcB  \\
Planet-to-star radius ratio & \rpl/\rstar         & $U(0,1)$          & \pA   & \pB \\
Impact parameter & $b$                 &  $U(0,1)$      & \bA   & \bB \\
Stellar reflex velocity semi-amplitude & $K$ \dotfill (m s$^{-1}$) & $U(0,1000)$   & \KA   & \KB   \\
Orbital eccentricity & $e$ & $F(0)$ & 0 (adopted) & 0 (adopted) \\
\noalign{\medskip}
\hline
\hline
Fitted parameter & Symbol \dotfill (Unit) & \multicolumn1c{Prior} & \multicolumn2c{\stnameA} \\
\hline
\hline
\noalign{\smallskip}
Stellar density & $\rho_\star$ \dotfill (kg m$^{-3}$)  & $LU(10^{2},10^{4})$ & \multicolumn2c{\rhostA} \\
Quad. limb-dark. coeff. (TESS-TRAPPIST) & $q_1^{\rm TESS-TRAPPIST}$ \dotfill & $U(0,1)$    & \multicolumn2c{\qaTESSTRAPPIST}\\
Quad. limb-dark. coeff. (TESS-TRAPPIST) & $q_2^{\rm TESS-TRAPPIST}$ \dotfill & $U(0,1)$    & \multicolumn2c{\qbTESSTRAPPIST}\\
Quadratic limb-darkening coefficient (NGTS) & $q_1^{\rm NGTS}$ \dotfill & $U(0,1)$    & \multicolumn2c{\qaNGTS}\\
Quadratic limb-darkening coefficient (NGTS) & $q_2^{\rm NGTS}$ \dotfill & $U(0,1)$    & \multicolumn2c{\qbNGTS}\\
Linear LD coeff. (LCO,KeplerCam,SAAO) & $q_1^{\rm LCO-KeplerCam}$ \dotfill & $U(0,1)$    & \multicolumn2c{\qaLCOKeplerCamSAAO}\\
Linear LD coeff. (MEarth,ULMT,PEST) & $q_1^{\rm ULMT-PEST}$ \dotfill & $U(0,1)$    & \multicolumn2c{\qaMEarthULMTPEST}\\
Photometric jitter (TESS sector 3) & $\sigma_w^{\rm TESS-S3}$ \dotfill (ppm) & $LU(10^{-3},10^{6})$ & \multicolumn2c{\sigmaTESSA} \\
Photometric jitter (NGTS) & $\sigma_w^{\rm NGTS}$ \dotfill (ppm) & $LU(10^{-3},10^{6})$ & \multicolumn2c{\sigmaNGTS} \\
Photometric jitter (TESS sector 30) & $\sigma_w^{\rm TESS-S30}$ \dotfill (ppm) & $LU(10^{-3},10^{6})$ & \multicolumn2c{\sigmaTESSB} \\
Photometric jitter (LCO) & $\sigma_w^{\rm LCO}$ \dotfill (ppm) & $LU(10^{-3},10^{6})$ & \multicolumn2c{\sigmaLCO} \\
Photometric jitter (MEarth) & $\sigma_w^{\rm MEarth}$ \dotfill (ppm) & $LU(10^{-3},10^{6})$ & \multicolumn2c{\sigmaMEarth} \\
Photometric jitter (KeplerCam) & $\sigma_w^{\rm KeplerCam}$ \dotfill (ppm) & $LU(10^{-3},10^{6})$ & \multicolumn2c{\sigmaKeplerCam} \\
Photometric jitter (ULMT) & $\sigma_w^{\rm ULMT}$ \dotfill (ppm) & $LU(10^{-3},10^{6})$ & \multicolumn2c{\sigmaULMT} \\
Photometric jitter (TRAPPIST) & $\sigma_w^{\rm TRAPPIST}$ \dotfill (ppm) & $LU(10^{-3},10^{6})$ & \multicolumn2c{\sigmaTRAPPIST} \\
Photometric jitter (PEST) & $\sigma_w^{\rm PEST}$ \dotfill (ppm) & $LU(10^{-3},10^{6})$ & \multicolumn2c{\sigmaPEST} \\
Photometric jitter (SPECULOOS) & $\sigma_w^{\rm SPECULOOS}$ \dotfill (ppm) & $LU(10^{-3},10^{6})$ & \multicolumn2c{\sigmaSPECULOOS} \\
Photometric jitter (SAAO) & $\sigma_w^{\rm SAAO}$ \dotfill (ppm) & $LU(10^{-3},10^{6})$ & \multicolumn2c{\sigmaSAAO} \\
Relative flux offset (TESS sector 3) & mflux$^{\rm TESS-S3}$ \dotfill (ppm) & $N(0.0,0.1)$ & \multicolumn2c{\mfluxTESSA} \\
Relative flux offset (NGTS) & mflux$^{\rm NGTS}$ \dotfill (ppm) & $N(0.0,0.1)$ & \multicolumn2c{\mfluxNGTS} \\
Relative flux offset (TESS sector 30) & mflux$^{\rm TESS-S30}$ \dotfill (ppm) & $N(0.0,0.1)$ & \multicolumn2c{\mfluxTESSB} \\
Relative flux offset (LCO) & mflux$^{\rm LCO}$ \dotfill (ppm) & $N(0.0,0.1)$ & \multicolumn2c{\mfluxLCO} \\
Relative flux offset (MEarth) & mflux$^{\rm MEarth}$ \dotfill (ppm) & $N(0.0,0.1)$ & \multicolumn2c{\mfluxMEarth} \\
Relative flux offset (KeplerCam) & mflux$^{\rm KeplerCam}$ \dotfill (ppm) & $N(0.0,0.1)$ & \multicolumn2c{\mfluxKeplerCam} \\
Relative flux offset (ULMT) & mflux$^{\rm ULMT}$ \dotfill (ppm) & $N(0.0,0.1)$ & \multicolumn2c{\mfluxULMT} \\
Relative flux offset (TRAPPIST) & mflux$^{\rm TRAPPIST}$ \dotfill (ppm) & $N(0.0,0.1)$ & \multicolumn2c{\mfluxTRAPPIST} \\
Relative flux offset (PEST) & mflux$^{\rm PEST}$ \dotfill (ppm) & $N(0.0,0.1)$ & \multicolumn2c{\mfluxPEST} \\
Relative flux offset (SPECULOOS) & mflux$^{\rm SPECULOOS}$ \dotfill (ppm) & $N(0.0,0.1)$ & \multicolumn2c{\mfluxSPECULOOS} \\
Systemic radial velocity (FEROS) & $\gamma_{\rm FEROS}$ \dotfill (m s$^{-1}$)  & $U(21\,300,21\,500)$ & \multicolumn2c{\muFEROSA} \\
Radial-velocity jitter (FEROS) & $\sigma_{\rm FEROS}$ \dotfill (m s$^{-1}$)& $LU(10^{-3},10^3)$ & \multicolumn2c{\sigmawFEROSA} \\
Systemic radial velocity (HARPS) & $\gamma_{\rm HARPS}$ \dotfill (m s$^{-1}$)  & $U(21\,350,21\,450)$ & \multicolumn2c{\muHARPSA} \\
Linear radial-velocity drift & $\dot\gamma$ \dotfill (m s$^{-1}$ d$^{-1}$)          & $U(-10,10)$   & \multicolumn2c{\dmuA} \\
Quadratic radial-velocity drift & $\ddot\gamma$ \dotfill (m s$^{-2}$  d$^{-1}$)          & $U(-10,10)$  & \multicolumn2c{\ddmuA} \\
\noalign{\medskip}
\hline
\hline
Derived parameter & Symbol \dotfill (Unit) & \dotfill & \plnameA & \plnameB \\
\hline
\hline
\noalign{\smallskip}
Planetary mass & \mpl\ \dotfill (\mearth)           & \dotfill & \mpAearth & \mpBearth \\
\dotfill & \mpl\ \dotfill (\mjup)           & \dotfill & \mpA & \mpB \\
Planetary radius & \rpl\ \dotfill (\rjup)           & \dotfill & \rpA & \rpB \\
Planetary density & \rhopl\ \dotfill (kg m$^{-3}$)   & \dotfill & \rhoA & \rhoB \\
Scaled semi-major axis & $a$/\rstar	\dotfill & \dotfill		& \arstarA	& \arstarB \\
Orbital semi-major axis & $a$ \dotfill (au)    & \dotfill & \smaA & \smaB \\
Orbital inclination & $i$ \dotfill (deg) & \dotfill  & \incA & \incB \\
Planetary equilibrium temperature ($A=0.5$) & \teq \dotfill (K)    & \dotfill & \tqA & \tqB \\ 
\hline
\enddata
\end{deluxetable*}

\begin{acknowledgments}
The NGTS facility is operated by the consortium institutes with support from the UK Science and Technology Facilities Council (STFC) under projects ST/M001962/1 and ST/S002642/1. 

This paper includes data collected with the TESS mission, obtained from the MAST data archive at the Space Telescope Science Institute (STScI). Funding for the TESS mission is provided by the NASA Explorer Program. STScI is operated by the Association of Universities for Research in Astronomy, Inc., under NASA contract NAS 5–26555. We acknowledge the use of public TESS data from pipelines at the TESS Science Office and at the TESS Science Processing Operations Center.
This research has made use of the Exoplanet Follow-up Observation Program website, which is operated by the California Institute of Technology, under contract with the National Aeronautics and Space Administration under the Exoplanet Exploration Program.

Based on observations made with ESO Telescopes at the La Silla Paranal Observatory under program IDs: 
0104.A-9007,
0104.A-9012,
0108.A-9008,
0109.A-9024,
0109.A-9025 and
0110.A-9035 for FEROS; 
and
0104.C-0413,
0104.C-0588,
105.20FX,
105.20G9,
106.216H and
106.21ER
for HARPS.

This work makes use of observations from the Las Cumbres Observatory global telescope (LCOGT) network.
Part of the LCOGT telescope time was granted by NOIRLab through the Mid-Scale Innovations Program (MSIP). MSIP is funded by NSF. 

This paper makes use of data from the MEarth Project, which is a collaboration between Harvard University and the Smithsonian Astrophysical Observatory. The MEarth Project acknowledges funding from the David and Lucile Packard Fellowship for Science and Engineering, the National Science Foundation under grants AST-0807690, AST-1109468, AST-1616624 and AST-1004488 (Alan T. Waterman Award), the National Aeronautics and Space Administration under Grant No. 80NSSC18K0476 issued through the XRP Program, and the John Templeton Foundation.

The University of Liège’s contribution to SPECULOOS has received funding from the European Research Council under the European Union’s Seventh Framework Programme (FP/2007-2013; grant Agreement no. 336480/SPECULOOS), from the Balzan Prize and Francqui Foundations, from the Belgian Scientific Research Foundation (F.R.S.- FNRS; grant no. T.0109.20), from the University of Liège, and from the ARC grant for Concerted Research Actions financed by the Wallonia-Brussels Federation. The Cambridge contribution is supported by a grant from the Simons Foundation (PI: Queloz, grant number 327127). The Birmingham contribution research is in part funded by the European Union’s Horizon 2020 research and innovation programme (grant’s agreement no. 803193/BEBOP), from the MERAC foundation, and from the Science and Technology Facilities Council (STFC; grant no. ST/S00193X/1).

This research uses observations made with the University of Louisville Manner Telescope (ULMT) at Mt. Lemmon Observatory. 

TRAPPIST-South is funded by the Belgian Fund for Scientific Research (Fond National de la Recherche Scientifique, FNRS) under the
grant FRFC 2.5.594.09.F, with the participation of the Swiss National
Science Fundation (SNF). 

This paper uses observations made at the South African Astronomical Observatory (SAAO).

Some of the observations in the paper made use of the High-Resolution Imaging instrument ‘Alopeke. ‘Alopeke was funded by the NASA Exoplanet Exploration Program and built at the NASA Ames Research Center by Steve B. Howell, Nic Scott, Elliott P. Horch, and Emmett Quigley. ‘Alopeke was mounted on the Gemini North telescope of the international Gemini Observatory, a program of NSF’s NOIRLab, which is managed by the Association of Universities for Research in Astronomy (AURA) under a cooperative agreement with the National Science Foundation. on behalf of the Gemini partnership: the National Science Foundation (United States), National Research Council (Canada), Agencia Nacional de Investigaci\'{o}n y Desarrollo (Chile), Ministerio de Ciencia, Tecnolog\'{i}a e Innovaci\'{o}n (Argentina), Minist\'{e}rio da Ci\^{e}ncia, Tecnologia, Inova\c{c}\~{o}es e Comunica\c{c}\~{o}es (Brazil), and Korea Astronomy and Space Science Institute (Republic of Korea).

Based in part on observations obtained at the Southern Astrophysical Research (SOAR) telescope, which is a joint project of the Minist\'{e}rio da Ci\^{e}ncia, Tecnologia e Inova\c{c}\~{o}es (MCTI/LNA) do Brasil, the US National Science Foundation’s NOIRLab, the University of North Carolina at Chapel Hill (UNC), and Michigan State University (MSU).

RB acknowledges support from FONDECYT project 11200751 and from ANID -- Millennium  Science  Initiative -- ICN12\_009.
The postdoctoral fellowship of KB is funded by F.R.S.-FNRS grant T.0109.20 and by the Francqui Foundation.
KAC acknowledges support from the TESS mission via subaward s3449 from MIT.
This publication benefits from the support of the French Community of Belgium in the context of the FRIA Doctoral Grant awarded to MT.
DJA is supported by UKRI through the STFC (ST/R00384X/1) and EPSRC (EP/X027562/1).
AC gratefully acknowledges support by the Heising-Simons Foundation (Grant No. 2020-1825)
EG gratefully acknowledges support from the UK Science and Technology Facilities Council (STFC; project reference ST/W001047/1).
JSJ gratefully acknowledges support by FONDECYT grant 1240738 and from the ANID BASAL project FB210003.
Some of this work has been carried out within the framework of the NCCR PlanetS supported by the Swiss National Science Foundation under grants 51NF40\_182901 and 51NF40\_205606.
\end{acknowledgments}

\facilities{
    ESO:3.6m~(HARPS),
    Max Planck:2.2m~(FEROS),
    LCOGT,
    MEarth,
    NGTS,
    TESS,
    ULMT,
    TRAPPIST.
}

\bibliography{ngts11c}
\bibliographystyle{aasjournalv7}

\end{document}